\documentclass[onecolumn,aps,prl,superscriptaddress,amsfonts]{revtex4-1}

\usepackage{physics}

\usepackage{amsmath}
\usepackage{fancyhdr}
\usepackage{amssymb}
\usepackage{amsthm}
\usepackage{graphicx}
\usepackage{caption}

\usepackage[
            bookmarks=true,pdfpagelabels=true,
            plainpages=false,
            pageanchor=true,
						]{hyperref}

\newcommand{\Yb}{$^{171}\textrm{Yb}^+$}
\newcommand{\Sstate}{^2\textrm{S}_{1/2}}
\newcommand{\Phalf}{^2\textrm{P}_{1/2}}
\newcommand{\upspin}{\ket{\uparrow}_z}
\newcommand{\downspin}{\ket{\downarrow}_z}

\newcommand{\LevelDiagram}{
\begin{figure}[t!]
\centering
\includegraphics*[width=0.37 \columnwidth]{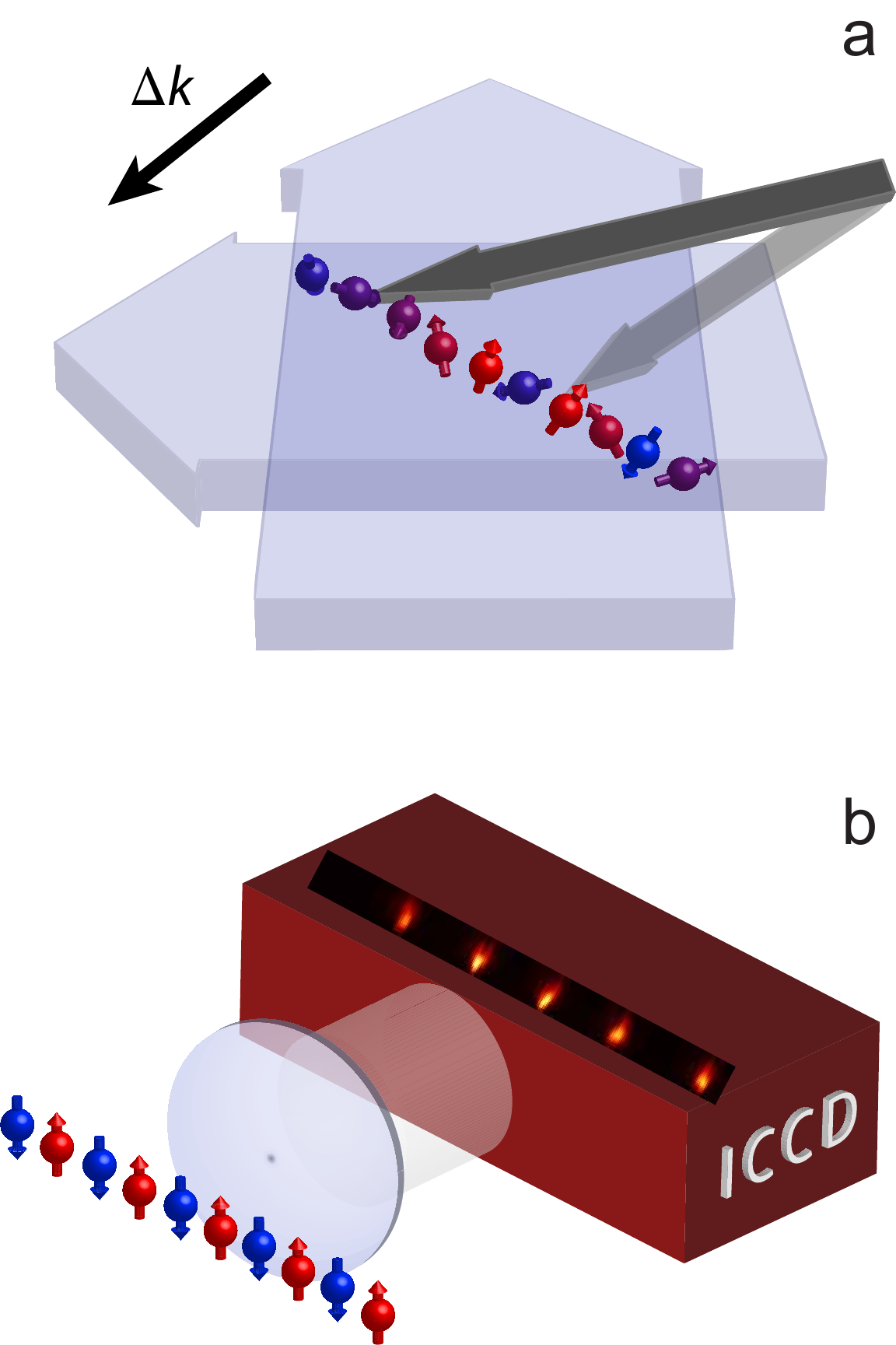}
\includegraphics*[width=0.5 \columnwidth]{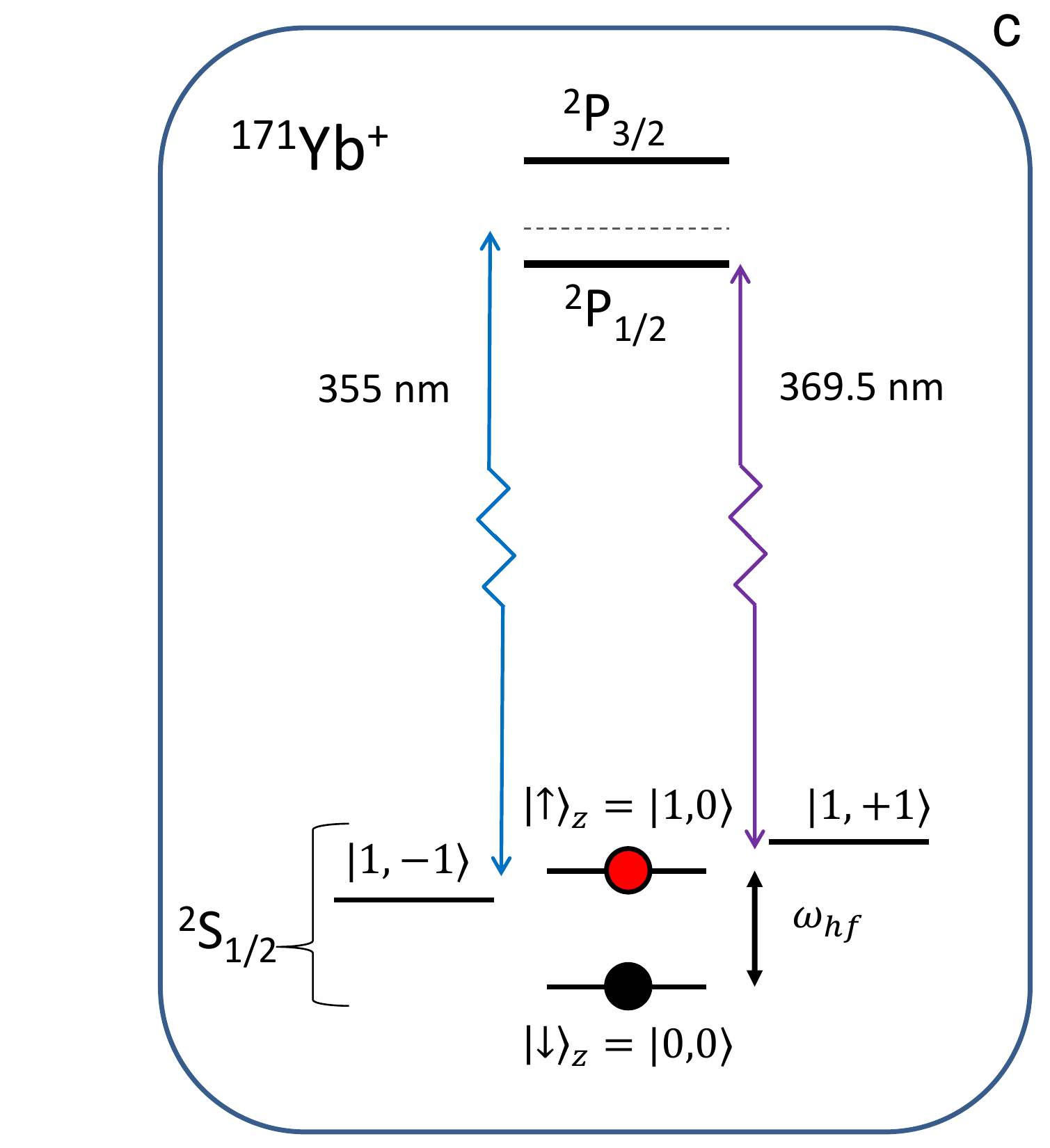}
\caption{\textbf{Experimental Schematic: (a)} A 1D chain of trapped ion spins evolves under the influence of global Raman beams (blue) which generate long-range transverse field Ising interactions. Local disorder can be realized by rastering a tightly focused individual addressing laser (grey) across selected ions. \textbf{(b)} Ion magnetizations $\expval{\sigma_i^z}$ are imaged through state dependent fluorescence on an intensified camera (ICCD). \textbf{(b)} Energy level diagram of \Yb. The ground state manifold has four sublevels, but the two picked for the qubit states $\upspin$ and $\downspin$ are magnetically insensitive `clock states'.}
\label{fig:LevelDiagram}
\end{figure}
}

\newcommand{\JMeasurementFigure}{
\begin{figure}[t!]
\includegraphics*[width=\columnwidth]{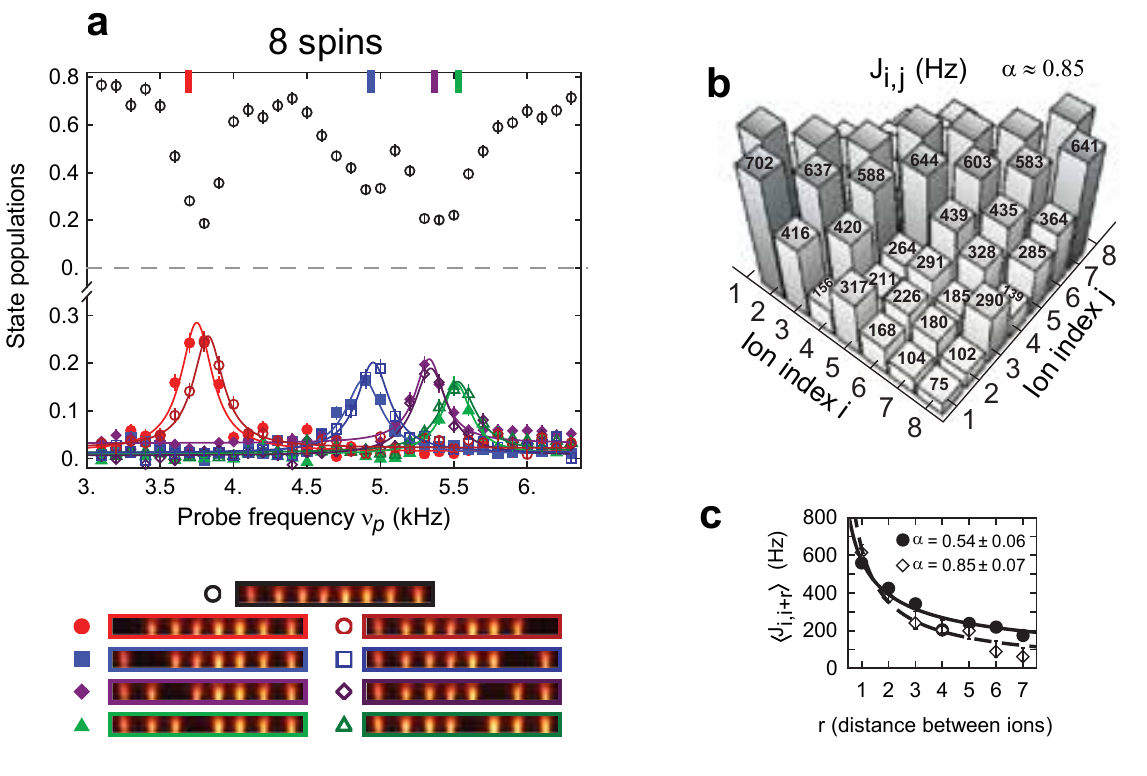}
\caption{\textbf{Spectroscopic Hamiltonian Measurement: } \textbf{(a)} Scanning the frequency of a weak, oscillating magnetic field probe $\nu_p$ will cause population transfer between an initially polarized state $\ket{\uparrow \uparrow \uparrow \uparrow \uparrow \uparrow \uparrow \uparrow}_x$ to those with a single spin flipped when the probe is resonant with their frequency difference. Depleted population in the initial state is observed simultaneously with an increased population in the excited states. \textbf{(b)} Measurement of $N(N-1)/2$ independent frequency splittings is sufficient to invert the many-body spectra and solve for the coupling matrix $J_{i,j}$. \textbf{(c)} The average measured interaction range is fit to a power law decay in two cases, with good agreement to a range calculated from the trap frequencies and laser beatnote detunings. Adapted from Ref.~\cite{Senko2014}.}
\label{fig:Spectroscopy}
\end{figure}
}

\newcommand{\MBLFigure}{
\begin{figure}[t!]
\includegraphics*[width=\columnwidth]{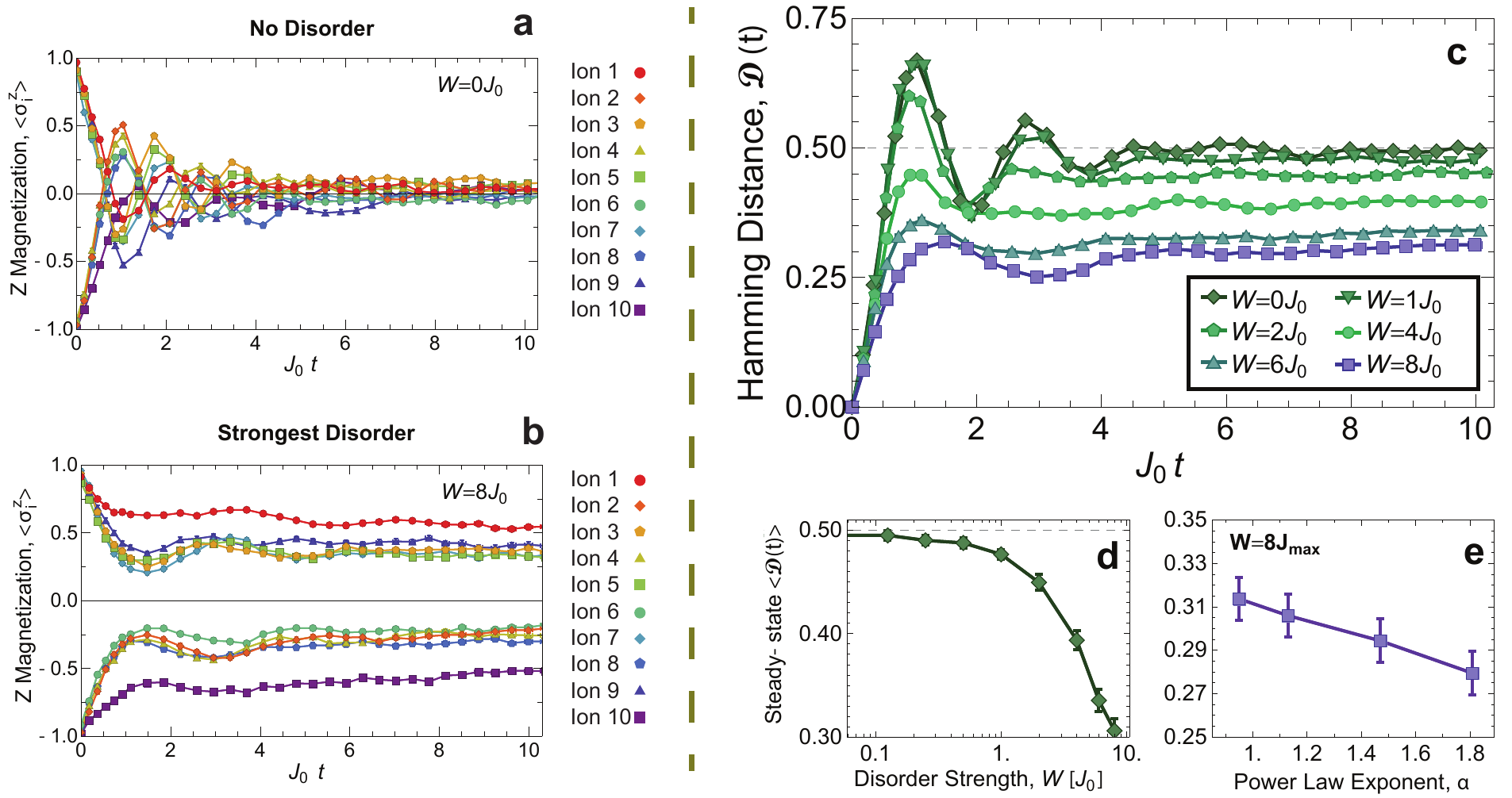}
\caption{\textbf{Many-Body Localization: } \textbf{(a)} With no disorder present, the prepared N\'{e}el state of 10 ions thermalizes to a state consistent with $\expval{\sigma_i^z} \approx 0$ for each spin. \textbf{(b)} With the strongest achievable disorder, $W = 8 J_0$, all spins retain memory of initial conditions with  $|\expval{\sigma_i^z}| \neq 0$. \textbf{(c)} The normalized Hamming distance $\mathcal{D}(t)$ has reached a steady state value for $J_0 t \geq 5$ at all measured disorder strengths. \textbf{(d)} The time averaged steady state value $\expval{\mathcal{D}(t)}$ after the plateau exhibits the onset of a crossover between a thermalizing regime $(\expval{\mathcal{D}(t)} = 0.5)$ and localizing regime $(\expval{\mathcal{D}(t)} = 0)$. \textbf{(e)} The steady state Hamming distance increases with longer range interactions, indicative of the movement away from single particle Anderson localization limit at $\alpha = \infty$. In all plots, each time series is an average over 30 instances of disorder. Adapted from Ref.~\cite{Smith2016}}.
\label{fig:MBL}
\end{figure}
}

\newcommand{\PrethermFigure}{
\begin{figure}[t!]
\includegraphics*[width=\columnwidth]{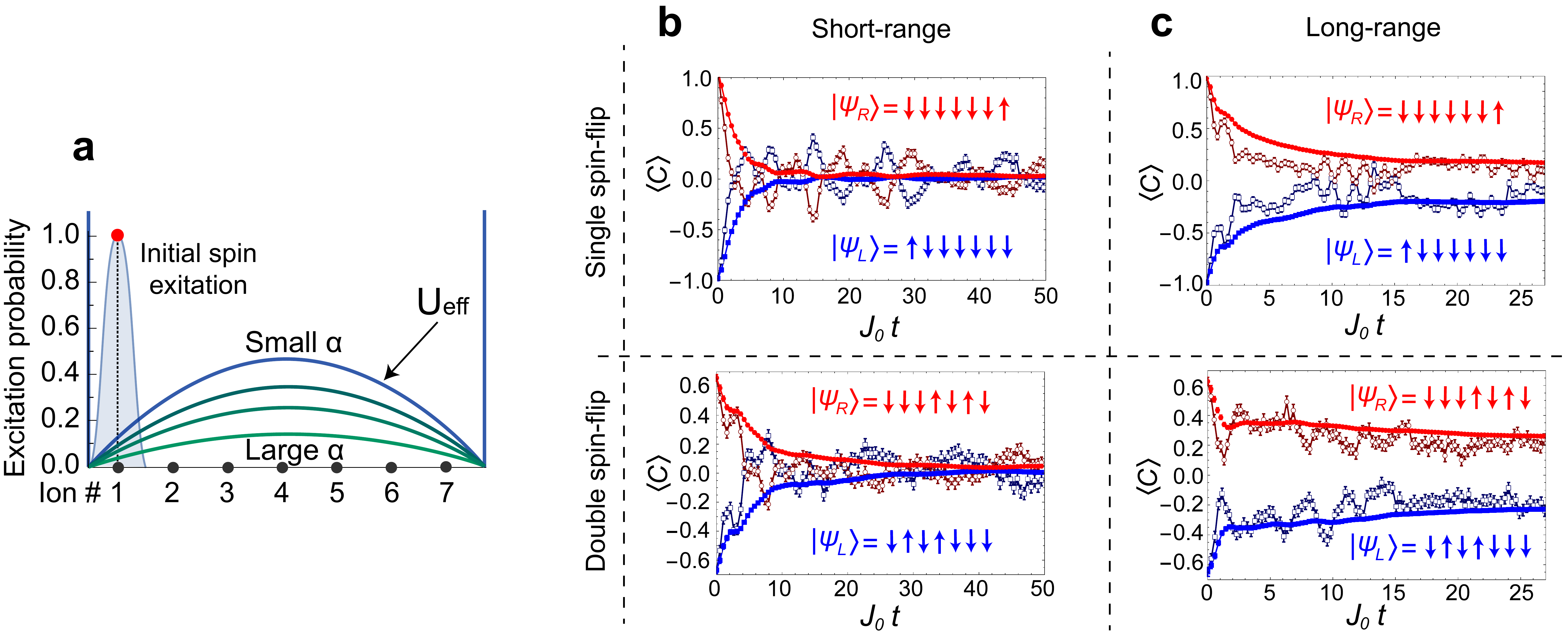}
\caption{\textbf{Prethermalization: } \textbf{(a)} An initial spin excitation is prepared on one side of a 7 ion chain subject to open boundary conditions and long-range $XY$ interactions. As the range increases ($\alpha$ decreases) the excitation is subject to an emergent potential barrier. \textbf{(b)} In the case of short range interactions, either one or two spin flips delocalize to $\expval{C} \approx 0$, consistent with the GGE. \textbf{(c)} For long-range interactions memory of initial conditions in preserved in a long-lived perthermal state. In both (b) and (c), the open squared/circles plot $\expval{C}$ for initial states prepared on the left/right side of the spin chain, while the filled circles/squares plot the cumulative time average $\expval{\overline{C}}$ for this data. Adapted from Ref.~\cite{Neyenhuis2016}.}
\label{fig:Pretherm}
\end{figure}
}

\newcommand{\DTCFigure}{
\begin{figure}[t!]
\includegraphics*[width=\columnwidth]{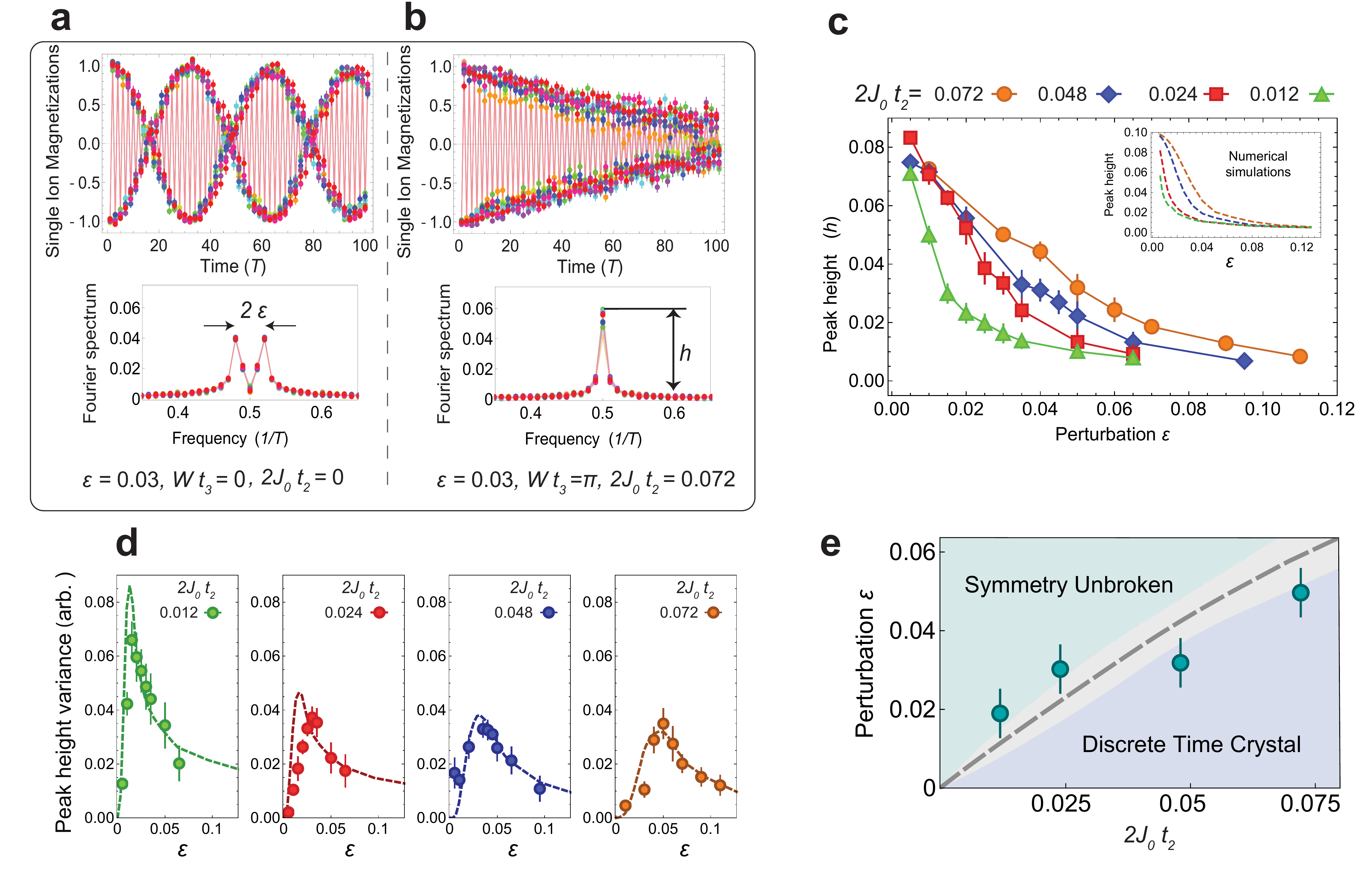}
\caption{\textbf{Discrete Time Crystals: }\textbf{(a)} Spin dynamics under repeated application of $H_1$. A perturbation that modifies the frequency response is clearly visible in the Fourier domain by two peaks separated by $2 \varepsilon$. \textbf{(b)} When the MBL terms $H_2$ and $H_3$ are also applied all spins oscillate in phase and with a common frequency of $1/(2T)$. \textbf{(c)} The height of the Fourier peak $h$ at $1/(2T)$ is used as an order parameter to demonstrate how larger $\varepsilon$ destroys the DTC order at four different values of $J_0$. \textbf{(d)} The variance of $h$ diverges at the phase transition boundary for each value of $J_0$, and exhibits good agreement with numerical predictions (dashed lines). \textbf{(e)} The critical $\varepsilon$ at the peak of the data in (d) demonstrates a linear dependence with $J_0 t_2$. This boundary separates DTC from symmetry unbroken phases in the thermodynamic limit. Adapted from Ref.~\cite{Zhang2017}.}
\label{fig:DTC}
\end{figure}
}

\begin{document}
\title{Non-thermalization in trapped atomic ion spin chains}



\author{P.~W.~Hess,$^{1}$ P.~Becker,$^{1}$ H.~B.~Kaplan,$^{1}$ A.~Kyprianidis}
\affiliation{Joint Quantum Institute, University of Maryland Department of Physics and National Institute of Standards and Technology, College Park, Maryland 20742, USA}
\author{A.~C.~Lee}
\affiliation{Joint Quantum Institute, University of Maryland Department of Physics and National Institute of Standards and Technology, College Park, Maryland 20742, USA}
\affiliation{Northrop Grumman Corporation, Linthicum, MD 21090}
\author{B.~Neyenhuis}
\affiliation{Joint Quantum Institute, University of Maryland Department of Physics and National Institute of Standards and Technology, College Park, Maryland 20742, USA}
\affiliation{Honeywell, Broomfield, CO 80021, USA}
\author{G.~Pagano}
\affiliation{Joint Quantum Institute, University of Maryland Department of Physics and National Institute of Standards and Technology, College Park, Maryland 20742, USA}\author{P.~Richerme}
\affiliation{Joint Quantum Institute, University of Maryland Department of Physics and National Institute of Standards and Technology, College Park, Maryland 20742, USA}
\affiliation{Department of Physics, Indiana University, Bloomington, Indiana 47405, USA}
\author{C.~Senko}
\affiliation{Joint Quantum Institute, University of Maryland Department of Physics and National Institute of Standards and Technology, College Park, Maryland 20742, USA}
\affiliation{Institute for Quantum Computing and Department of Physics \& Astronomy, University of Waterloo, Waterloo, Ontario N2L 3G1, Canada}
\author{J.~Smith}
\affiliation{Joint Quantum Institute, University of Maryland Department of Physics and National Institute of Standards and Technology, College Park, Maryland 20742, USA}
\affiliation{Institute for Defense Analyses, Alexandria, VA 22311}
\author{W.~L.~Tan,$^{1}$ J.~Zhang,$^{1}$ C.~Monroe}
\affiliation{Joint Quantum Institute, University of Maryland Department of Physics and National Institute of Standards and Technology, College Park, Maryland 20742, USA}




\date{\today}

\begin{abstract}
Linear arrays of trapped and laser cooled atomic ions are a versatile platform for studying emergent phenomena in strongly-interacting many-body systems. Effective spins are encoded in long-lived electronic levels of each ion and made to interact through laser mediated optical dipole forces. The advantages of experiments with cold trapped ions, including high spatiotemporal resolution, decoupling from the external environment, and control over the system Hamiltonian, are used to measure quantum effects not always accessible in natural condensed matter samples. In this review we highlight recent work using trapped ions to explore a variety of non-ergodic phenomena in long-range interacting spin-models which are heralded by memory of out-of-equilibrium initial conditions. We observe long-lived memory in static magnetizations for quenched many-body localization and prethermalization, while memory is preserved in the periodic oscillations of a driven discrete time crystal state.   


\end{abstract}



\maketitle

\section{Introduction}
It is well known that highly complex, non-linear, and chaotic classical systems will typically reach a thermal equilibrium state~\cite{Feynman1998}. It is natural to ask how generic this phenomenon might be and whether it can be applied to quantum systems as well. Of particular interest is the extent to which thermalization occurs in closed quantum systems, where many of the requirements for classical ergodicity do not apply~\cite{Cazalilla2010, Polkovnikov2011, DAlessio2016}. These questions have led to an extensive body of theoretical work for understanding and classifying the differences between thermalizing and non-thermalizing quantum systems. Much of this is based on the eigenstate thermalization hypothesis~\cite{Deutsch1991, Srednicki1994, Rigol2008}, which predicts that generic observables in strongly-interacting quantum systems exhibit thermal properties in their massively entangled eigenstates. There are notable exceptions, like many-body localization (MBL), where systems fail to thermalize despite strong interactions between constituent particles~\cite{Basko2007, Oganesyan2007, Nandkishore2015}. New analytic and numerical techniques have been developed~\cite{Polkovnikov2011, White1992, Kamenev2009, Verstraete2008} to make predictions about these highly excited and out of equilibrium quantum states, where the typical tools of statistical physics do not apply. 

As with any field of scientific inquiry, experimental measurements are required to confirm these predictions and improve the theoretical tools. Experimental studies using synthetic quantum matter made of trapped and cooled atoms have demonstrated remarkable versatility to study many of these engineered non-ergodic theories~\cite{Kinoshita2006,  Billy2008, Gring2012, Schreiber2015, Kondov2015, Kaufman2016}. These experiments benefit from a high degree of isolation from thermal environments, making them excellent models for closed quantum systems. A broad range of tunable Hamiltonians can be generated using externally applied electromagnetic radiation. Often these optical or microwave radiation fields can be extinguished rapidly, allowing access to highly excited but coherent quantum state dynamics via quench or Floquet processes.

This review will focus on recent experiments which have been performed with linear crystals of trapped atomic ions. Such systems can be used to realize long-range transverse field Ising models (TFIM) in 1D by encoding an effective spin in the internal states of each ion. It is possible to access both chaotic and non-ergodic regimes by controlling interaction strength and range, applying axial or transverse magnetic fields, and generating local magnetic field disorder. In section~\ref{sec:Tools} we begin with an overview of experimental tools for generating all relevant terms in these long-range Ising models. In sections~\ref{sec:MBL} and \ref{sec:Pretherm}, we discuss observations of many-body localization in the presence of disorder and prethermalization of spin excitations in clean long-range systems. Both are examples of long-lived memory of initial conditions, a hallmark of non-ergodicity. Finally in section~\ref{sec:DTC} we discuss the observation of a discrete time crystal, a new non-equilibrium phase of matter with robust and long-lived temporal periodicity, followed by future directions in section~\ref{sec:Conclusion}.

\section{Tools for Quantum Simulation with Trapped Ions}
\label{sec:Tools}
High addressability~\cite{Lee2016}, long range entanglement~\cite{Monz2011, Debnath2016}, and high fidelity readout techniques~\cite{Noek2013} make trapped atomic ions a versatile platform for both quantum computation~\cite{Monroe2013} and simulation~\cite{Blatt2012}. Effective spins are encoded in internal electronic levels in each trapped atom. The Coulomb force strongly couples the ions' motional states and provides the mechanism for long-range spin-spin coupling via laser mediated spin-phonon interactions~\cite{Molmer1999, Porras2004}. 

In particular, the experiments outlined in this review were performed on chains of \Yb~in linear radiofrequency (Paul) traps~\cite{Major2005}. Effective spins are encoded in the $\Sstate$ hyperfine ground states of \Yb, which we will label $\ket{\uparrow}_z \equiv \ket{F= 1, m_F = 0}$ and $\ket{\downarrow}_z \equiv \ket{F=0, m_F = 0}$~\cite{Olmschenk2007}. A laser tuned to 369.5~nm strongly couples the ground $\Sstate$ and excited $\Phalf$ states and is used for state initialization and readout (Figure~\ref{fig:LevelDiagram}c). Trapped ion motion is cooled into the Lamb-Dicke regime by laser cooling, and the spins can be polarized using an optical pumping transition. Imaging the ions with spin-dependent fluorescence allows high fidelity readout with short exposure times~\cite{Noek2013} (Figure \ref{fig:LevelDiagram}b). 

\LevelDiagram

\subsection{Generating Long-Range Transverse Field Ising Interactions}
We perform coherent operations on the \Yb~qubits using stimulated Raman transitions. Two overlapped laser beams at 355~nm provide a momentum transfer $\Delta k$ along the transverse direction of the ion chain (Figure~\ref{fig:LevelDiagram}a). If the beatnote between them is set to the qubit frequency ($\omega_{hf}/2\pi = 12.642819$~GHz), then the spins are coherently rotated with a two-photon Rabi frequency $\Omega \sim 1$~MHz. Resolved sideband interactions between spin and motion result if the beatnote is detuned from the spin transition frequency by an amount $\mu$ near the motional frequencies $\omega_m$, provided we are in the Lamb-Dicke regime $\eta = \Delta k \sqrt{\hbar/2 M \omega_m} \ll 1$, for ion mass $M$.

Effective spin-spin interactions are engineered with an off resonantly driven optical dipole force~\cite{Molmer1999}. Stimulated Raman transitions with a bichromatic beatnote detuning $\pm \mu$ generate an evolution operator
\begin{equation}
U(t) = \exp\left[\sum_i \hat{\phi}_i(t) \sigma_i^x + i \sum_{i,j} \chi_{i,j}(t) \sigma_i^x \sigma_j^x\right].
\end{equation}
Here $\hat{\phi}_i(t)$ describes spin-dependent displacements of phononic excitations in phase space, while $\chi_{i,j}(t)$ is a phonon independent spin-spin coupling term~\cite{Zhu2006, Kim2009}. 

We operate in the detuned regime where $|\mu - \omega_m| \gg \eta \Omega$, which keeps $\hat{\phi}_i(t)$ small and bounded. The state evolution is described by an effective Ising Hamiltonian in spin space alone (with $\hbar = 1$)
\begin{equation}
H_{SS} = \sum_{i<j}J_{i,j} \sigma_i^x \sigma_j^x.
\end{equation}
The coupling matrix $J_{i,j}$ is given by a sum over the ion couplings to all normal modes
\begin{equation}
J_{i,j} = \Omega^2 \omega_R \sum_{m=1}^N \frac{b_{i,m} b_{j,m}}{\mu^2-\omega_m^2},
\end{equation}
where $\omega_R = \hbar \Delta k^2/2M$ is the atomic recoil frequency and $b_{i,m}$ is the normalized eigenmode matrix for motional mode $\omega_m$.

In the experiments described here, the spin-spin interactions are generated by ``global'' Raman lasers which produce equal intensity and beatnote detuning ($\delta_m = \mu - \omega_m$) on each ion. The spin model realized under these conditions takes the form of a power law decaying with distance
\begin{equation}
J_{i,j} \approx \frac{J_0}{|i-j|^\alpha},
\end{equation}
where $J_0$ is the average nearest neighbor strength and $\alpha$ the tunable range. The range can be adjusted continuously from an all-to-all coupling of $\alpha = 0$ when the contribution from the center-of-mass mode dominates ($\delta_{COM} \ll \delta_m$) to a short-range dipolar interaction of $\alpha = 3$ as $\delta_m \rightarrow \infty$. 

For these experiments, the stimulated Raman transitions are driven by a mode-locked tripled $\textrm{Nd:YVO}_4$ laser at 355~nm. The 120~MHz repetition rate and $\sim 14$~ps pulse width of this laser creates a frequency comb with enough bandwidth to span the qubit hyperfine splitting~\cite{Hayes2010}. We tune $J_{i,j}$ by adjusting $\mu$ or the trap voltages and are limited to a range of $(0.5 < \alpha < 2)$ and $J_0/2 \pi \leq 1$~kHz by the available laser power and the Paul trap geometry. Tunability of $\mu$ and the Rabi frequency ($\Omega(t)$) is provided by an acousto-optic modulator (AOM) driven by an arbitrary waveform generator (AWG), and drifts in the laser's repetition rate are accounted for by stabilizing the beatnote at $\omega_{hf}$ using a feedback scheme~\cite{Islam2014}. 

Both transverse and axial field Ising models can be realized by adding effective magnetic fields 
\begin{equation}
H_B = B/2 \sum_{i} \sigma_i^\gamma
\end{equation}
where $\gamma = \left(x,y,z\right)$. Magnetic fields in $\gamma = \left(x,y\right)$ are generated by driving stimulated Raman transitions resonant with the qubit frequency $\omega_{hf}$, where the phase of this third frequency component with respect to the bichromatic beatnotes determines the axis of rotation in the $xy$-plane. Alternatively, by applying a global shift to the bichromatic beatnotes by an amount $B$, a rotating frame shift between the qubit and the beatnotes generates an effective field in $\gamma = z$~\cite{Lee2016a}. In either case, we limit $B/2 \ll \eta \Omega \ll \delta_{COM}$ in order to prevent unwanted higher order terms in our effective Hamiltonian, practially limiting us to $B/2\pi \leq 10$~kHz.

\JMeasurementFigure

The particular shape of $J_{i,j}$ and its dependence on $\delta_m$ can be verified using a spectroscopic technique~\cite{Senko2014}. We weakly modulate the transverse magnetic field such that
\begin{equation}
H=\sum_{i<j}J_{ij} \sigma_i^x \sigma_j^x + (B_0 + B_p \sin (2 \pi \nu_p t) \sum_{i} \sigma_i^y.
\end{equation}
When the probe frequency $\nu_p$ is resonant with an energy splitting $\Delta E = |E_a - E_b|$ between eigenstates $\ket{a}$ and $\ket{b}$, the transverse field will drive transitions if there is a nonzero matrix element $\bra{b} \sum_i \sigma_i^y \ket{a} \neq 0$. In the case of a weak transverse field, $(B_0, B_p) \ll J_0$, this corresponds to transitions between the eigenstates of $\sum_i \sigma_i^x$ that have a single spin flipped (eg $\ket{\uparrow \uparrow \ldots \uparrow}_x$ and $\ket{\downarrow \uparrow \ldots \uparrow}_x$). By scanning the probe frequency and monitoring the population transfer between eigenstates, we can directly measure the eigenstate energy splittings of our many-body Hamiltonian (Figure~\ref{fig:Spectroscopy}a). 

For an eigenstate transition that has flipped spin $i$, the measured energy difference is $\Delta E_i = 2 \sum_{j} J_{i,j}$. Measuring $N(N-1)/2$ energy splittings provides sufficient information to solve this system of equations and extract each component of $J_{i,j}$ (Figure~\ref{fig:Spectroscopy}b,c). We can also utilize individual addressing to measure components of $J_{i,j}$ by shelving all ions except $i, j$ into dark Zeeman sublevels $(\ket{1,\pm 1})$ that do not interact when $H_{SS}$ is applied~\cite{Smith2016, Jurcevic2014}. The oscillation frequency between $\ket{\downarrow_i \downarrow_j}_z \rightarrow \ket{\uparrow_i \uparrow_j}_z$ is a direct measurement of $J_{i,j}$. Both techniques have similar scaling with system size to measure $J_{i,j}$ with arbitrary couplings, but the spectroscopic method avoids the additional experimental overhead required for individual addressing. Such a procedure will prove especially useful for verifying more complex $J_{i,j}$ matrices, which can be engineered with individually addressed stimulated Raman interactions~\cite{Korenblit2012}.



%

\subsection{Individual Addressing for State Preparation and Programmable Disorder}
A critical tool for realizing a non-ergodic MBL Hamiltonian is the ability to apply controlled disorder to the system. In this case, we apply local effective magnetic fields of the form
\begin{equation}
H_D = \sum_{i} D_i \sigma_i^z.
\end{equation}
These interactions are generated with a tightly focused 355~nm laser beam which imposes a light shift $2D_i$ on the bare qubit frequency. For \Yb~qubits the differential shift from the (2nd order) AC Stark effect between the $\upspin$ and $\downspin$ states from 355~nm light is highly suppressed~\cite{Campbell2010}. This is advantageous for maintaining a stable qubit with long coherence time, but makes it difficult to apply disorder using an interaction like $H_D$. 


When the light field contains coherent frequency components with a frequency difference ($\delta_2$) not far detuned from the qubit frequency, a fourth-order light shift can become significant and dominate over the typical second-order (AC Stark) shifts~\cite{Lee2016, Mizrahi2013}. When $\delta_2 = 0$, this light field drives stimulated Raman transitions with Rabi frequency $\Omega = g_0^2 / 2 \Delta$, where $g_0$ is the resonant Rabi frequency of the $6\textrm{S} \rightarrow 6\textrm{P}$ transition and $\Delta$ is the detuning of the 355~nm laser from the $\Phalf$ states. The resulting effective interaction can shift the qubit frequency by
\begin{equation}
\delta \omega^{(4)} = \frac{|\Omega|^2}{2 \delta_2}.
\end{equation}
The fourth order shift $\delta \omega^{(4)}$ has a quadratic dependence on the laser intensity since $g_0^2 \propto I$. The relative strength of the 4th order and 2nd order light shifts is $\delta \omega^{(4)}/\delta \omega^{(2)} = \frac{g_0^2}{4 \delta_2 \omega_{hf}}$, which means the 4th order shift dominates in the limit of high laser intensity $g_0^2 \gg \delta_2 \omega_{hf}$. 



In our case since these frequency components arise from a mode-locked laser with 120~MHz repetition rate, $|\delta_2| \leq 60$~MHz. With a tight focus the intensity is sufficiently high that we observe a clear quadratic dependence of the qubit shift on laser intensity~\cite{Lee2016}. The 355~nm light for individual addressing is focused through our imaging objective lens to achieve a beam waist of a few microns. This limits crosstalk between individually addressed ions to $< 2\%$ and provides sufficiently high fidelity single qubit light shifts for our application. 

An acousto-optic deflector (AOD) is used to translate the lateral position of the addressing beam focus and select which ion to address. Although multiple ions can be addressed simultaneously by driving the AOD with multiple frequencies, this leads to an inefficient use of our optical power because of the quadratic intensity dependence of the fourth order light shift. A linear scaling of $\delta \omega^{(4)}$ per ion can be recovered by instead rastering across all ions to be addressed. In the limit that the raster frequency is much faster than other relevant frequency scales in our quantum simulation, then this simply acts like the desired time averaged local magnetic field with $D_i = \expval{\delta \omega^{(4)}_i} / 2$.

This site dependent shift can also be utilized for preparation of arbitrary initial product states~\cite{Neyenhuis2016}. A Ramsey scheme provides the highest fidelity technique for flipping individual spins. The spins are optically pumped then rotated into $\ket{\downarrow}_x$, at which time $H_D$ is applied with equal light shifts $D$ to select ions. The light shifted ions will precess faster and after time $t = \pi / (2 D)$ they will be $180^{\circ}$ out of phase from the other spins. A final $\pi/2$ pulse then rotates this into an arbitrary spin eigenstate in the $z-$basis. 


\section{Many-Body Localization in a Trapped Ion Spin Chain}
\label{sec:MBL}
With the above described experimental tools, we can study various dynamical phenomena in the transverse field Ising model, which exhibit rich quantum phases and transitions between them. Many-body localization is an intense field of interest, whereby disorder gives rise to non-ergodic behavior even in strongly interacting systems~\cite{Basko2006, Oganesyan2007, Nandkishore2015}. This effect is a generalization of single-particle ``Anderson'' localization~\cite{Anderson1958}, which is well understood theoretically and has been measured in a number of experimental platforms~\cite{Wiersma1997, Billy2008, Roati2008, Kondov2011, Schwartz2007, Dalichaouch1991, Hu2008}. In Anderson localization, experiments are limited to regimes of low excitation energy and no interparticle interactions. In the case of MBL, non-ergodicity and lack of thermalization apply even at much higher initial energy densities and temperatures, along with a broad set of interaction ranges and disorder strengths. 

Although a number of observables have been identified to characterize phase transitions between MBL and ergodic states~\cite{Luitz2015}, the main experimental signature probed in trapped ion spin chains is the long-lived memory of initial conditions~\cite{Hauke2015, Iyer2013}. We begin the experiment by preparing the 10 spin N\'{e}el state with staggered order $(\ket{\psi_0} = \ket{\downarrow \uparrow \downarrow \uparrow \downarrow \uparrow \downarrow \uparrow \downarrow \uparrow}_z)$ which is highly excited with respect to the disordered Ising Hamiltonian
\begin{equation}
H_{MBL} = \sum_{i<j} J_{i,j} \sigma_i^x \sigma_j^x + B/2 \sum_{i} \sigma_i^z + \sum_{i} D_i \sigma_i^z
\end{equation}
where $D_i$ is sampled from a uniform distribution, $D_i \in [-W/2 , W/2]$, where $W$ is the width of the disorder distribution. This Hamiltonian is rapidly quenched on and the resulting single spin magnetization dynamics $\expval{\sigma_i^z(t)}$ are measured for times up to $t = 10/J_0$. 

\MBLFigure

In the absence of disorder these same initial spin states will thermalize if the uniform transverse field $B$ is sufficiently large~\cite{Deutsch1991, Srednicki1994, Rigol2008}. We prepare eigenstates of both $\sigma^x$ and $\sigma^z$ and measure the resulting single ion magnetization projected into those directions. In the case of a thermalizing system we expect information of generic initial conditions to be lost in all directions of the Bloch sphere, namely $\expval{\sigma_i^x} = \expval{\sigma_i^z} = 0$. We observe that the system is thermalizing at short times ($J_0 t < 1$) for values of transverse field $B \geq 4 J_0$ (Figure~\ref{fig:MBL}a). 

The transverse field is held fixed at $B = 4 J_0$ in order to study how disorder localizes the spin (Figure~\ref{fig:MBL}b). Each measurement of magnetization dynamics for disorder width $W$ is repeated with 30 different realizations of disorder, which are subsequently averaged together. This number of realizations is sufficient to reduce the disorder sampling error to be smaller than the features of interest. Averaging over initial states is unnecessary because MBL applies over a broad range of initial energies which are sampled by the disorder averaging.

We observe after some initial growth and oscillations, the magnetization of each spin settles to a steady state value for $J_0 t \geq 5$. To quantify the degree of localization, we compute the normalized Hamming distance (HD)
\begin{equation}
	\begin{aligned}
		\mathcal{D}(t) &=  \frac{1}{2} - \frac{1}{2N} \sum_{i} \expval{\sigma_i^z(t)\sigma_i^z(0)}{\psi_0} \\
               &=  \frac{1}{2} - \frac{1}{2N} \sum_{i} (-1)^i  \expval{\sigma_i^z(t)}.
	\end{aligned}
\end{equation}
This counts the number of spin flips from the initial state, normalized to the length of the chain. At long times, a randomly oriented thermal state shows $\mathcal{D} = 0.5$ while one that remains fully localized has $\mathcal{D} = 0$ (Figure~\ref{fig:MBL}c). The average steady state value $\mathcal{D}(t)$ for $J_0 t \geq 5$ serves as an order parameter for probing the crossover between localizing and thermalizing parameter regimes. 

As a function of $W$, the disorder clearly pushes the spin chain towards a localized regime (Figure~\ref{fig:MBL}d). Likewise, the localization strengthens as $\alpha$ is increased towards shorter range interactions (Figure~\ref{fig:MBL}e), recovering Anderson localization via a Jordan-Wigner transformation in the $\alpha \rightarrow \infty$ limit. Numerical studies have confirmed that full localization occurs within experimentally accessible disorder strengths and interaction ranges~\cite{Wu2016}. A many-body delocalization transition predicted at $\alpha > 1.5$ is absent from our data, and is likely due to finite size effects~\cite{Burin2015a}.  Characterizing and eliminating these finite size effects is a major goal of future planned experiments using much longer chains of trapped \Yb~ions.

Many-body localization is a unique case in which a closed quantum system remains non-ergodic and localized even up to infinite times. Such a prediction is hard to verify experimentally, as there is always some some finite coupling to the environment which will thermalize the system over accessible time scales~\cite{Bordia2016, Luschen2017}. This makes it difficult to distinguish truly MBL phases from those in which the dynamics are glassy or metastable, but will thermalize over long timescales even in a closed system~\cite{Reimann2008}.

\section{Prethermalization - Memory without disorder}
\label{sec:Pretherm}
While the disordered Hamiltonians necessary for MBL are inherently non-integrable, it is also interesting to study the degree to which integrability breaking can lead to non-ergodic behavior in clean systems~\cite{Polkovnikov2011, Larson2013, Manmana2007}. The long-range transverse field Ising models realized in our trapped ion quantum simulator present a unique context to study this since the degree of integrability breaking can be tuned via the interaction range~\cite{Kiendl2016}. Such near integrable systems typically show metastable plateaus where memory of initial conditions is preserved for long, but not infinite, times.


Prethermalization describes the physics of these metastable states and the mechanisms by which they stay out of equilibrium for extended periods of time~\cite{Berges2004}. Much of the experimental realizations of prethermalization involve nearly integrable systems, such as 1D Bose gases~\cite{Kinoshita2006, Gring2012, Langen2015}, which can still be characterized by approximately conserved quantities associated with integrability. The prethermalization in such cases is accurately described by relaxation to a distribution given by a generalized Gibbs ensemble (GGE)~\cite{Rigol2007}.

In contrast, the long-range interactions realized in our trapped ion quantum simulator create a prethermal state which is sufficiently far from integrability that it can not be predicted by a GGE~\cite{Neyenhuis2016, Gong2013, Marcuzzi2013}. We analyze this prethermalization using a Holstein-Primakoff transformation, where spin excitations in an Ising chain are mapped to free bosons. For nearest neighbor interactions with open boundary conditions, these bosons propagate in a square well potential. An increased interaction range raises a potential barrier in the middle of the spin chain, resulting in prethermal confinement of spin-excitations created on either side of the barrier (Figure~\ref{fig:Pretherm}a). The long-time memory of initial conditions, relaxed by tunneling through this emergent barrier, is the observed signature of this prethermal state. 


\PrethermFigure

The experiment begins by preparing a single spin excitation on either edge of a 7-ion chain $\ket{\psi_{R}} = \ket{\downarrow \downarrow \downarrow \downarrow \downarrow \downarrow \uparrow}_z$ or $\ket{\psi_{L}} = \ket{\uparrow \downarrow \downarrow \downarrow \downarrow \downarrow \downarrow}_z$. We then quench on the long range TFIM Hamiltonian
\begin{equation}
H = \sum_{i<j} J_{i,j} \sigma_i^x \sigma_j^x + B \sum_{i} \sigma_i^z
\end{equation}
in the limit where $B \gg J_0$ and for interaction ranges of either $\alpha = 0.55$ (long-range) or $\alpha = 1.33$ (short-range). In this limit, the strong $B$ field makes it energetically forbidden to create or destroy spin excitations, and the initial spin excitation will propagate ballistically through the spin chain~\cite{Jurcevic2014, Richerme2014}. 

We quantify the position of the spin excitation by measuring the observable
\begin{equation}
C(t) = \sum_{i} \frac{2i-N-1}{N-1} \frac{\sigma_i^z +1}{2}
\end{equation}
which ranges from $-1$ to $1$ if the spin excitation is on the left/right side of the chain, respectively. Memory of initial conditions is observed only in the case of long-range interactions, where $\textrm{sign}(\expval{C(t)}) =  \textrm{sign}(\expval{C(0)})$ for times up to $J_0 t = 25$, regardless of initial conditions. The effect is pronounced in the cumulative time average $\expval{\overline{C}}$, which averages out high frequency oscillations. The observed prethermalization is robust to weak interactions between spin excitations. When two spin excitations are prepared in $\ket{\psi_0} = \ket{\downarrow \uparrow \downarrow \uparrow \downarrow \downarrow \downarrow}$, which is even further from integrability than the single spin flip case, the resulting dynamics are still prethermal in the presence of long-range interactions (Figure~\ref{fig:Pretherm}b,c).


This state should persist in the thermodynamic limit, where the long-range interactions and open boundary conditions continue to effectively break the translational invariance in the bulk. We observed that this effect still persisted when we more than tripled the length of the spin chain to 22 ions. 
This experiment serves as a good example of non-ergodicity breaking in clean systems and is a clear demonstration that prethermal behavior cannot always be described using the formalism of the GGE.

\section{Discrete Time Crystals}
\label{sec:DTC}
So far we have considered cases where a quantum quench to a static Hamiltonian leads to non-ergodic quantum dynamics. Time dependent or periodic Hamiltonians also support out-of-equilibrium phases with spin dynamics that are either robustly oscillatory and localized in frequency space, or dephase from ergodic evolution~\cite{Ponte2015, DAlessio2014}. Take the example of the quantum kicked rotor, where the periodic (Floquet) Hamiltonian generates dynamics which can be tuned between chaotic and temporally localizing regimes~\cite{Chirikov1981, Fishman1982}. Many of the robustly oscillatory phases of matter in Floquet systems have no direct analogs to phenomena in quenched static Hamiltonians~\cite{Khemani2016, Potter2016}. 

A discrete (Floquet) time crystal is an example of a non-ergodic time dependent phase, where symmetry breaking spin oscillations show robustness to variations in the system Hamiltonian~\cite{Sacha2015, Khemani2016, Else2016a, Yao2017}. This effect has a close connection to many-body localization, which is used here as a mechanism to stabilize the time crystal dynamics and prevent heating due to the Floquet drive~\cite{DAlessio2014, Lazarides2014, Ponte2015}. While a static MBL Hamiltonian causes long lived memory of initial conditions in spin magnetizations, a Floquet MBL Hamiltonian preserves the memory of the initial phase of oscillatory dynamics.

\DTCFigure

In the case of a discrete time crystal (DTC) this long lived oscillation exhibits a form of discrete time translational symmetry breaking. A discrete symmetry is imposed on the system by the periodicity $T$ of the Hamiltonian, $H(t) = H(t+T)$. However, the DTC state spontaneously breaks this symmetry by oscillating at a sub-harmonic frequency with period $2T$. When an MBL Hamiltonian is part of the Floquet drive cycle, the symmetry breaking is robust to perturbations in the drive. Both the symmetry breaking oscillations and their robustness from many-body interactions provide an analogy to spontaneous breaking of spatial symmetry in the formation of solid crystals~\cite{Wilczek2013}. The original proposals for observing time crystals in the spontaneous breaking of continuous time translational symmetry in a ground state were shown to be impossible~\cite{Bruno2013b, Nozieres2013, Watanabe2015}, but the DTC is a generalization of this concept.

We perform an experiment driving a 10 ion spin chain with a periodic Hamiltonian to realize the DTC~\cite{Zhang2017}. We prepare an initial state polarized in $\ket{\psi_0} = \ket{\downarrow}_x$, then apply a three component Floquet Hamiltonian with overall period $T = t_1 + t_2 +t_3$
\begin{equation}
H = \begin{cases} 
H_1 = g(1-\varepsilon) \sum_i \sigma^y_i, & \mbox{time } t_1  \\[4pt]  
H_2 = \sum_i J_{i,j} \sigma^x_i \sigma^x_{j}, \ &\mbox{time } t_2  \\[4pt]  
H_3 = \sum_{i}{D_i}\sigma_i^x \ & \mbox{time } t_3.
\end{cases}
\end{equation}
where $g$ is the effective Rabi frequency tuned such that $2 g t_1 = \pi$ and $\varepsilon$ is a fractional perturbation varied between $0 - 15\%$. We observe the magnetization $\expval{\sigma_i^x}$ after each Floquet period $T$ and evolve the system out to 100 periods. The data is analyzed in the frequency domain by applying a discrete Fourier transform to the time series, where the expected sub-harmonic oscillations will occur at a frequency of $\nu_{tc} = 1/(2T)$.

The first term $H_1$ is a perturbed $\pi$ pulse which breaks the system's time translational symmetry. In the absence of $H_2$ and $H_3$, the spin magnetizations do not generically oscillate at the sub-harmonic frequency, and the system instead tracks the drive at frequency $\nu = T^{-1} (1/2 - \varepsilon)$ (Figure~\ref{fig:DTC}a). Terms $H_2$ and $H_3$ collectively serve to generate a Hamiltonian deep in the MBL regime.\footnote{These two terms are temporally separated because additional $\pi/2$ pulses are necessary to rotate the light shift generated $\sigma_i^z$ disorder into the $x-$direction~\cite{Lee2016}. The Floquet evolution is equivalent to their simultaneous application because $\comm{H_2}{H_3} = 0$.} The short range $(\alpha = 1.5)$ interactions are weak $J_0 t_2 < 0.04$ compared to the disorder, which is pulled from a uniform distribution of width $W$ set to $W t_3 = \pi$.  

The application of $H_2$ and $H_3$ with small $\varepsilon$ restores the magnetization to oscillations at sub-harmonic frequency $\nu_{tc}$ (Figure~\ref{fig:DTC}b). To quantify the degree of temporal ordering, the amplitude of the Fourier transform at the sub-harmonic frequency $h(\nu_{tc})$ is used as an order parameter (Figure~\ref{fig:DTC}c). By varying $\varepsilon$ and $J_0$, we can probe the cross over from DTC to symmetry unbroken phases and observe that the temporal order is destroyed when $\varepsilon \gtrsim 2 J_0 t_2$. The variance of $h(\nu_{tc})$, taken across 10 disorder instances, peaks at the cross over boundary as would be indicative of a phase transition in the thermodynamic limit (Figure~\ref{fig:DTC}d). The peak of the variance observable exhibits a linear relationship between $\varepsilon$ and $J_0$ as expected, revealing the parameter regime where the DTC is stable against perturbations (Figure~\ref{fig:DTC}e). 

Discrete time crystals are not specific to 1D spin chains and have also been observed in a 3D crystal of nitrogen vacancy center spins in diamond~\cite{Choi2017}. These experiments point to the role that Floquet drives can play in generating entirely new phases of matter, which may not exist in static or equilibrium systems. These states can exhibit a gamut of features like symmetry protected topological order or non-ergodicity. Understanding all of these regimes will require new theoretical tools for treating highly out of equilibrium quantum systems, and complementary experiments on large and strongly interacting systems to verify them. 

\section{Conclusion and Future Directions}
\label{sec:Conclusion}
We have discussed a few contexts where chains of trapped ions, tailored with optical forces to realize strongly interacting spin models, can be made to exhibit non-ergodic many-body quantum dynamics~\cite{Smith2016, Neyenhuis2016, Zhang2017}. There is promise in expanding these studies to observe new examples of non-ergodic phenomena in quench or Floquet type experiments. Performing similar experiments in 2D ion crystals would help answer questions about the viability of MBL in more than one dimension~\cite{DeRoeck2016}. Penning traps have been used to confine large rotating 2D ion crystals with tunable long range interactions~\cite{Britton2012, Bohnet2016}, but similar crystals created in Paul traps would benefit from improved spatiotemporal resolution for spin initialization and readout~\cite{Yoshimura2015, Richerme2016}.



While 2D Ising models lead to new phenomenon through enchanced spin connectivity, it is also possible to explore how higher dimensional spins $S > 1/2$ would effect the observed examples of non-ergodicity. For example, the DTC has modified sub-harmonic ordering for $S=1$ systems~\cite{Yao2017, Choi2017}, and there might be many other unique effects considering the topological nature of integer spin Heisenberg chains~\cite{Haldane1983a, Affleck1988}. Integer spin dynamics have also been demonstrated with our trapped \Yb~atoms~\cite{Senko2015, Cohen2014, Cohen2015}, where three spin states can be encoded if we include additional Zeeman sublevels in the $\Sstate$ manifold: $\ket{+} = \ket{F=1, m_F=1}$, $\ket{-} = \ket{F=1, m_F=-1}$, and $\ket{0} = \ket{F=0, m_F=0}$. We can generate an effective Hamiltonian by applying a bichromatic field with beat frequencies $\omega_- + \mu$ and $\omega_+ - \mu$, where $\omega_{\pm}$ is the frequency splitting between the $\ket{0}$ and $\ket{\pm}$ states. The Hamiltonian in this case is an $XY$-interaction between $S=1$ particles
\begin{equation}
H_{eff} = \sum_{i<j} \frac{J_{i,j}}{4} (S_i^+ S_j^- + S_i^- S_j^+) + \sum_{i,m} V_{i,m} \left[S_i^z - (S_i^z)^2\right]
\end{equation}
where $S_i^\pm$ are raising and lowering operators and the matrix $V_{i,m}$ is a local field term proportional to the phononic excitations in the ion chain. These local field terms can be eliminated by addition of two more beat-note frequencies at $\omega_- - \mu$ and $\omega_+ + \mu$, which generate a long-range Ising interaction $H=\sum_{i<j}J_{i,j} S_i^x S_j^x$ from a generalization of the M\o lmer S\o rensen scheme used for simulating $S=1/2$ chains~\cite{Molmer1999}. 

We have observed both coherent quench dynamics between two interacting $S=1$ ions and adiabatic state preparation in chains of 2-4 spins by ramping down a global $(S^z)^2$ term~\cite{Senko2015}. Using our individual addressing laser, we could generate controllable terms of the form $\alpha S_i^z + \beta (S_i^z)^2$ in order to engineer a disordered $S=1$ chain.  This provides a toolbox for $S=1$ quantum simulations as complete as that for $S=1/2$, and opens the door to experiments studying localization in integer spin chains~\cite{Grass2013, Strinati2016}. 


So far all the experiments described here have been performed on relatively small system sizes of up to 22 spins. Most of the physics here can be verified using exact or approximate numerical techniques, including exact-diagonalization or the density-matrix-renormalization-group (DMRG). This has provided an excellent opportunity to demonstrate the capability of trapped ion quantum simulation, while benchmarking their performance on these small systems. However, once the system sizes are increased beyond $\sim 30$ spins, exact calculations become impossible on highly excited and entangled states like MBL. In this case larger experiments, with well controlled interactions that scale efficiently with system size are necessary for verifying the numerical calculations and observing new phenomenon in large condensed-matter-like systems. 

Efforts are underway to dramatically increase the number of ions in a linear crystal and achieve these goals. A current limitation in our experiments is the residual background gas molecules in the vacuum chamber, which can collide with and destroy our long and fragile trapped ion chain. Our solution was to engineer a cryogenically pumped ion trap chamber, which promises to have the background pressure and collision rate reduced by orders of magnitude. We have demonstrated that we can trap up to 120 ions in a linear configuration for hours at a time, an important step towards simulating long-range spin models in large systems. By combining this larger system with the techniques for Hamiltonian control developed thus far, we hope to push into the regime of ``quantum supremacy'', where our experiments can help understand complex many-body systems in ways inaccessible to current numerical techniques.




\section{Acknowledgements}
PWH drafted the manuscript. All authors edited and approved the manuscript. C. Monroe is the co-founder and Chief Scientist of IonQ, Inc. All other authors declare that they have no competing interests. This work is supported by the ARO Atomic Physics Program, the AFOSR MURI on Quantum Measurement and Verification, and the NSF Physics Frontier Center at JQI.


\bibliographystyle{naturemag-mod-Hess}
\bibliography{Hess_References_3_2017}




\end{document}